\documentclass[useAMS,usenatbib]{mn2e} 
\usepackage[paper=letterpaper,margin=0.75in]{geometry}
\usepackage{graphicx}
\usepackage{epsfig}
\usepackage{amsmath}

\newcommand{\hs}{\hspace{1mm}} 
\newcommand{\apj}{ApJ}
\newcommand{\aap}{A\&A} 
\newcommand{\apjl}{ApJL}
\newcommand{\mnras}{MNRAS} 

\newcommand{\apjs}{ApJS} 
\newcommand{\nat}{{\it Nature}}
 
\newcommand{\pasj}{PASJ}

\def\lsim{~\rlap{$<$}{\lower 1.0ex\hbox{$\sim$}}}

\def\gsim{~\rlap{$>$}{\lower 1.0ex\hbox{$\sim$}}}

\title[Are Theorists Modeling the Right LAEs?]{An Empirical Study of the Relationship between Ly$\alpha$ and UV selected Galaxies: Do Theorists and Observers `Select' the Same Objects?}

\author[Dijkstra \& Wyithe]{Mark
Dijkstra$^{1}$\thanks{E-mail:dijkstra@mpa-garching.mpg.de} and J. Stuart
B. Wyithe$^{2}$\\ $^{1}$Max Planck Institute for Astrophysics, Karl-Schwarzschild-Str. 1, 85741, Garching, Germany\\ $^{2}$School of Physics,
University of Melbourne, Parkville, Victoria, 3010, Australia}

\begin{document}

\date{\today} \pagerange{\pageref{firstpage}--\pageref{lastpage}}
\pubyear{2009}

\maketitle

\label{firstpage}
\begin{abstract}
Lyman Alpha Emitters (LAEs) are galaxies that have been selected on the basis of a strong Ly$\alpha$ emission line in their spectra. Observational campaigns over the last decade have dramatically increased the sample of known LAEs, which now extends out to $z=7$. These discoveries have motivated numerous theoretical studies on the subject, which usually define LAEs in their models based on sharp Ly$\alpha$ luminosity and equivalent width (EW) cuts. While broadly representative, this procedure does not mimic the selection from observational programs in detail, which instead use cuts in various colour-spaces. In this paper we investigate what implications this disjoint may have for theoretical studies that aim to model the observed population of LAEs. We construct an empirical model for the number density of star forming galaxies as a function of their UV and Ly$\alpha$ luminosity, utilising measured constraints on the luminosity functions of drop-out galaxies, and their luminosity dependent probability distribution function of Ly$\alpha$ EW. In particular, we investigate whether the LAE luminosity functions can be reproduced by defining LAEs using a ($z$-dependent) Ly$\alpha$ luminosity and EW threshold. While we are able to reproduce the observed distribution of Ly$\alpha$ EW among LAEs out to restframe EW$\sim 200$ \AA, we find that our formalism {\it over-predicts} both the UV and Ly$\alpha$ luminosity functions of LAEs by a factor of 2-3, and is inconsistent with observations at the $\sim 95$\% level. This tension is partially resolved if we assume the Ly$\alpha$ EW-distribution of drop-out galaxies to be truncated at restframe EW$\gsim 150$ \AA. However the overprediction indicates that modeling LAEs with simple REW and luminosity cuts does not accurately mimic observed selection criteria, and can therefore lead to uncertainties in the predicted number density of LAEs. On the other hand, the predicted redshift evolution is not affected. We apply our formalism to drop-out galaxies at $z>6$, and predict the luminosity functions of LAEs at $z=7-9$.
 \end{abstract}
 
\section{Introduction}
\label{sec:intro}

Two complementary  observational techniques have been very succesful in finding high-redshift galaxies. The Lyman Break technique -- or drop-out technique -- has been used to constrain the observed rest-frame UV luminosity functions of Lyman Break galaxies (LBGs) out to redshifts as high as $z=10$ \citep[e.g.][]{Bouwens10a,Bunker10,Finkelstein10,Yan10,Oesch11}. The Lyman Break technique relies on the fact that the spectra of star forming galaxies are strongly suppressed blueward of the Lyman-$\alpha$ resonance at $\lambda\lsim 1216$ \AA\hs \citep[][]{Steidel96}. The presence of this `break' in the spectrum is a direct consequence of the temperature of stellar atmospheres, and of the interstellar and intergalactic absorption of ionizing and Lyman series photons. The resulting break in the spectrum strongly suppresses the broad-band flux blueward of some--redshift dependent--filter that is observed from a galaxy (i.e. V drop-out galaxies at $z\sim 5$ are detected only in filters redder than V). 

Narrowband surveys have constrained luminosity functions of Ly$\alpha$ emitting galaxies (a.k.a Ly$\alpha$ emitters, or LAEs) out to $z\sim 9$ \citep[][]{Hu02,Iye,Ouchi08,Ouchi10,Willis08,Hibon10,Tilvi10,clement11}, while Lehnert et al. (2010) have also reported discovery of a $Y_{105}$ drop-out galaxy with strong Ly$\alpha$ emission. The narrow band  technique relies on the presence of a strong Ly$\alpha$ emission line, which mostly originates in galactic HII regions \citep[][]{PP67}. The presence of such a line can produce an excess of observed flux in a narrowband filter (FWHM$\sim 100$ \AA) that is larger than expected based on the observed flux in overlapping broadband filters \citep[e.g.][]{Rhoads00}. 

Existing observations have given us accurate determinations of the luminosity functions of LBGs and LAEs. The observed rest-frame UV luminosity function of drop-out galaxies decreases monotonically with redshift at $z\geq 3$ \citep[e.g.][]{Reddy09,Bouwens06,Bouwens07,Bouwens08}. 
These luminosity functions can be converted into a cosmic star formation rate density, which drops by more than an order of magnitude between $z=3$ and $z>6$ \citep[e.g.][]{Hopkins06,UVslope,Robertson10}. Furthermore, the broad band colours of drop-out galaxies become bluer with redshift, which indicates that star forming galaxies become increasingly dust-free \citep[][]{Stanway05,Bouwens10b,Finkelstein10}. 

The redshift evolution of the luminosity function of LAEs is different. The Ly$\alpha$ luminosity functions of LAEs are observed to be  remarkably constant between $z=3$ and $z=6$ \citep[e.g.][]{Hu98,Ouchi08}, after which the number density decreases at $z\ga6$ \citep[][but also see Hu et al. 2010]{Ka06,Ota08,Ota10,Ka11}. The first of these observations may be another consequence of the decreasing dust content of star forming galaxies towards higher redshift \citep[][]{Hayes10,Blanc11}. The second observation has received significantly more attention because it may signpost the existence of large regions of intervening neutral intergalactic gas, which are opaque to the Ly$\alpha$ photons \citep[][]{HS99,MR06,Ka06}. However, a proper interpretation of the redshift evolution of LAE luminosity functions at all redshifts requires understanding the detailed radiative transfer of Ly$\alpha$ photons through both the interstellar medium \citep[][]{Santos04,McK,DW10,Barnes11,D11}, and ionized intergalactic medium \citep[e.g.][also see e.g. Fernandez \& Komatsu 2008]{IGM,McK,Iliev08,ZZ1,Dayal10,Laursen10,D11}.

LAEs are selected differently in theoretical models and observations.  Observationally, LAEs are defined by their location in several 2-dimensional color-colour spaces. For example, Ouchi et al. (2008) define $z=3.1$ LAEs by requiring that $V-{\rm NB}503>1.2$ and [($V<V_{2\sigma}$ and $B-V>0.5$) or ($V\geq V_{2\sigma}$ and $B-V_{2\sigma}>0.5$), where $V_{2\sigma}=27.7$ denotes the 2$\sigma$ limiting magnitude of the V-band images]. 
The first requirement corresponds to having an excess flux in the narrowband filter denoted by NB503, while the broad-band selection corresponds to the requirement of having break in the spectrum blueward of the Ly$\alpha$ resonance. The requirement that the object have a strong narrowband excess translates {\it approximately} to a minimum restframe equivalent width (REW$_{\rm min}$) of the Ly$\alpha$ line. In detail REW$_{\rm min}$ depends on the S/N at which an object is detected (see e.g. Gronwall et al. 2007 for an extended discussion on this and other complications regarding the flux dependence of the survey volume). Representative values of REW$_{\rm min}$ are often reported in papers discussing measurements of the LAE luminosity function, but these numbers should be interpreted with caution (e.g. Ouchi et al. 2008). However, when describing the LAE population in numerical or semi-analytic calculations, theoretical models generally literally adopt the quoted minimum values of REW$_{\rm min}$ to select LAEs from model galaxies \citep[e.g.][]{LD06,LF,Mao07,McK,Ko10,N10,Dayal10,ZZ1,Shimizu11,Forero}. In addition, some theoretical papers only apply flux thresholds, which formally separates them further from observations.

In this paper, we investigate the impact of simplified selection criteria (i.e. joint REW and luminosity cuts) on the `predicted' number density of LAEs. In particular, we generate phenomenological models for the number density of star forming galaxies as a function of UV and Ly$\alpha$ luminosity. We then investigate whether we can reproduce the LAE data if we define LAEs by a (redshift dependent) Ly$\alpha$ luminosity and equivalent width (EW) threshold. Our empirical models are taken from observed drop-out luminosity functions at $3 \leq z \leq 6$,  combined with observational constraints on the prominence of Ly$\alpha$ emission lines in drop-out galaxies.  The Ly$\alpha$ equivalent width distribution for LBGs has been measured at $z=3$ by \citet{Shapley03}, and at $3 < z < 7$ for a slightly smaller sample by Stark et al. (2010, 2011). We then investigate whether it is possible to reproduce the observed Ly$\alpha$ luminosity functions (i.e. the number density of LAEs as a function of Ly$\alpha$ luminosity and UV magnitude) and equivalent width distributions, based solely on our knowledge of this LBG population. If this is not possible, then the analysis implies that utilising simplified selection criteria can have a strong impact on the predicted number of LAEs. 
Indeed, our results indicate that as models of LAEs mature, selection criteria that resemble those applied on the actual data will increasingly have to be taken into consideration. 

The outline of our paper is as follows: in \S~\ref{sec:form} we describe our formalism for `predicting' the Ly$\alpha$ luminosity function, and its uncertainties. In \S~\ref{sec:results} we present our results for Ly$\alpha$ luminosity functions at $z\lsim 7$, and provide a discussion of uncertainties in \S~\ref{sec:dis}. We also make predictions for LAE luminosity functions out to $z=10$ in \S~\ref{sec:z9}. Finally, in \S~\ref{sec:conc} we present our conclusions.  The
cosmological parameter values used throughout our discussion are
$(\Omega_m,\Omega_{\Lambda},\Omega_b,h)=(0.27,0.73,0.046,0.70)$
\citep{Komatsu08}.

\section{Empirical Relationship Between LBGs and LAEs}

\label{sec:form}

The range of UV-magnitudes probed by samples of LBG and LAEs\footnote{For many LAEs the UV continuum is also detected.} currently overlap. Furthermore, recent Ly$\alpha$ radiative transfer modeling shows that Ly$\alpha$ photons escape anisotropically from young, dusty, simulated galaxies \citep[][]{Laursen09}. This implies that the observed narrow-band excess of a star forming galaxy -- and hence, whether it would be selected into a sample of LAEs -- depends on its orientation relative to the observer. Furthermore, modelling of the observed spectral profiles of LBGs near the Ly$\alpha$ line suggests that LBGs observed not to have any Ly$\alpha$ emission are not a separate class of galaxy. Rather, these are also intrinsically strong Ly$\alpha$ emitters, in which radiative transfer effects transform the Ly$\alpha$ emission line into an absorption feature \citep{SV08,Atek09,Des10}\footnote{How frequently such a `transformation' occurs can depend on the metallicity of the interstellar gas \citep[e.g.][]{F11}.}. Together these observations indicate that the LBG and LAE populations are intrinsically the same, and are only separated as a consequence of Ly$\alpha$ radiative transfer through opaque (dusty) media. 

The possibility that Ly$\alpha$ radiative transfer plays an important role in separating LAEs from LBGs is underlined by the observation of low surface brightness, extended Ly$\alpha$ halos around LBGs \citep{Steidel11}.  \citet{Steidel11} argue that the Ly$\alpha$ radiation in these halos was emitted as nebular emission, which is then scattered to the observer in the (outflowing) circumgalactic medium. When the  flux in these halos is properly accounted for, the total observed Ly$\alpha$ flux places most LBGs in the LAE category. 

Thus, current evidence suggests that the LBG and LAE populations should be considered within the same theoretical framework. However, before this can be reliably pursued it is important to understand the observational selection criteria that define LAEs.

\subsection{The Formalism}

Our goal is to connect the observed numbers of LAEs and LBGs. In this section we therefore begin by introducing the formalism used in this paper. The number density of LAEs with Ly$\alpha$ luminosities in the range $L_{\alpha} \pm dL_{\alpha}/2$ is
\begin{eqnarray}
\nonumber
\Phi(L_{\alpha})dL_{\alpha}&=&dL_{\alpha}\times F\times (1+\delta_V)\times\\ 
 &&\hspace{0mm}\int_{M_{\rm min}}^{M_{\rm max}}dM_{\rm uv}\phi(M_{\rm uv})P(L_{\alpha}|M_{\rm uv}),
\label{eq:phi}
\end{eqnarray} where $\phi(M_{\rm uv})dM_{\rm UV}$ denotes the number density of galaxies with absolute AB-magnitude range $M_{\rm uv}\pm dM_{\rm UV}$, and $P(L_{\alpha}|M_{\rm uv})$ is the conditional probability density function (PDF) for $L_{\alpha}$ for a given $M_{\rm UV}$. The parameter $F$ is a normalisation factor which is discussed in \S~\ref{secF}. The factor $(1+\delta_V)\equiv \Delta_V$ accounts for cosmic variance, and we assume that $\Delta_V$ is drawn from a Gaussian distribution with a standard deviation $\sigma_V=0.36$ (estimated following the procedure of Somerville et al. 2004, see Moster et al. 2011 for an update of this work)\footnote{The cosmic variance recipe given by Somerville et al (2004) formally applies to spherical volumes. The narrowband survey of Ouchi et al. (2008) probes volumes that are close to cubical, and therefore Somerville et al. (2004) should provide reasonable estimates for their cosmic variance. For the appropriate survey volume of $0.5-0.9 \times 10^6$ Mpc$^{-3}$, Figure~3 of Somerville et al. (2004) gives $\sigma_{\rm DM}=0.03-0.06$. To get a conservatively large estimate for $\sigma$, we adopt a linear bias parameter of $b=6$ which is on the high end of the observed range \citep[e.g.][]{Shima03,Gawiser07,Kovac07,G10}. This gives us $\sigma \sim 0.18-0.36$.}. The integral is taken over the range $M_{\rm min}=-30.0$ to $M_{\rm max}=-12.0$. 

\subsubsection{The Drop-Out Luminosity Function $\phi(M_{\rm uv})dM_{\rm UV}$}
\label{sec:dropLF}

We assume that the LBG luminosity function $\phi(M{\rm uv})dM_{\rm uv}$ is described by a Schechter function, the parameters of which ($\phi^*,M^*_{\rm UV},\alpha$) we take from the literature. Table~\ref{table:param} summarizes the redshift dependence of our adopted parameters, and the references from which these were taken. Note that formally, the Ly$\alpha$ luminosity functions have been determined at $z=3.1$, $z=3.7$ and $z=5.7$. We have interpolated the UV luminosity functions to these same redshifts by assuming that $M^*_{\rm UV}$ evolves as $M^*_{\rm UV}=-21.02+0.36(z-3.8)$ \citep{Bouwens08} while keeping the other parameters fixed.

\begin{table}
\begin{minipage}{8cm}
\centering
\caption{Adopted parameters for $\phi(M_{\rm uv})$.}
\begin{tabular}{l c c c}
\hline\hline 
redshift & $\phi^*$ ($10^{-3}$ cMpc$^{-3}$)& $M^*_{\rm UV}$ &$\alpha$ \\
$z=3$\footnote{From Reddy \& Steidel (2009).} & $1.7\pm 0.5$& $-21.0\pm 0.1$ & $-1.73\pm 0.13$\\
$z=4$\footnote{From Bouwens et al. (2007).} & $1.3\pm 0.2$& $-21.0\pm 0.1$ & $-1.73\pm 0.05$\\
$z=6^b$ & $1.4^{+0.6}_{-0.4}$ & $-20.2\pm 0.2$& $-1.74\pm 0.16$\\
\hline

$z=7$\footnote{From Bouwens et al. (2008). The slope $\alpha$ was kept fixed at the value that was inferred from the lower redshift observations. To obtain the lower limit on $M^*_{\rm UV}$, the value for $\phi^*$ at $z=9$ was assumed to be the same as at redshift $7$.} & $1.1^{+1.7}_{-0.7}$&  $-19.8\pm 0.4$& $-1.74$ \\
$z=8$\footnote{From Bouwens et al. (2010a). The constraint on $M^*_{\rm UV}$ was obtained by assuming no evolution in  $\phi^*$ and $\alpha$.} & $1.1^{+1.7}_{-0.7}$& $-19.45$& $-1.74$\\
$z=9^c$ &$1.1^{+1.7}_{-0.7}$ & $\gsim -19.6$& $-1.74$\\
\hline\hline
\end{tabular}
\label{table:param}
\end{minipage}
\end{table}
\begin{table}
\begin{minipage}{8cm}
\centering
\caption{Parameters related to detection thresholds in the narrowband surveys.}
\begin{tabular}{l c c }
\hline
redshift & REW$_{\rm min}$\footnote{Taken from Table~3 of Ouchi et al. (2008).} & $L_{\alpha,{\rm min/max}}$ ($\frac{{\rm erg}}{{\rm s}}$)\footnote{The Ly$\alpha$ luminosity functions of Ouchi et al. (2008) has bins with a width of 0.2 dex (in log$_{10} L_{\alpha}$). To estimate $L_{\alpha,{\rm min}}$ ($L_{\rm max}$), we subtracted (added) 0.1 dex from (to) the Ly$\alpha$ luminosity of the faintest (brightest) bin.} \\
\hline\hline
$z=3.1$ & 64 \AA & 2 / 50$\times 10^{42}$\\
$z=3.7$ & 44 \AA   & 4 / 40$\times 10^{42}$\\
$z=5.7$& 27 \AA  & 2.5 / 40$\times 10^{42}$\\
\hline\hline
\end{tabular}
\label{table:obs}
\end{minipage}
\end{table}

\subsubsection{The Conditional Probability $P(L_{\alpha}|M_{\rm uv}$)}
\label{sec:plamuv}

The absolute AB UV-magnitude relates to the UV luminosity density $L_{{\rm UV},\nu}$ (in erg s$^{-1}$ Hz$^{-1}$) as $M_{\rm UV}=-2.5 \log L_{{\rm UV},\nu}+51.6$ \citep{Ouchi08}.
Because Ly$\alpha$ luminosity is simply the product of rest frame equivalent width (REW) and luminosity density (L$_{\lambda}$ in erg s$^{-1}$ \AA$^{-1}$), we can express Ly$\alpha$ luminosity as a function of REW and $L_{{\rm UV},\nu}$ as $L_{\alpha}=C\times {\rm REW}\times L_{{\rm UV},\nu}$. The constant $C\equiv \frac{\nu_{\alpha}}{\lambda_{\alpha}}\big{(}\frac{\lambda_{\rm UV}}{\lambda_{\alpha}} \big{)}^{-\beta-2}$, in which $\nu_{\alpha}=2.47 \times 10^{15}$ Hz, $\lambda_{\alpha}=1216\hs$\AA, $\beta\equiv  d\log L_{\lambda}/d\log \lambda$ (i.e. $L_{\lambda} \propto \lambda^{\beta}$ and $L_{\nu}\propto \nu^{-\beta-2}$), and $\lambda_{\rm UV}=1700$ \AA\hs denotes the restframe wavelength at which the UV continuum flux density was measured \citep[][]{DW09}. Throughout this work we assume $\beta=-1.7$ (see \S~\ref{sec:dispar}). We can thus recast the conditional probability $P(L_{\alpha}|M_{\rm uv})$ as a function of the REW-PDF as 

\begin{eqnarray}
P(L_{\alpha}|M_{\rm uv})=
\left\{ \begin{array}{ll}
     \ P(x|M_{\rm UV})\frac{\partial {\rm REW} }{ \partial L_{\alpha}}& x \in (x_{\rm min},x_{\rm max});\\
         \ 0 & \mbox{otherwise},\end{array} 
\right. 
\label{eq:pm1}
\end{eqnarray} 
where $x\equiv$REW=$L_{\alpha}/[C\times L_{{\rm UV},\nu}]$, and $x_{\rm min}$ denotes the minimum equivalent width REW$_{\rm min}$ that a given narrowband survey is sensitive to (for smaller values of REW the narrow band excess would be too small for the galaxy to make it into the sample of LAEs). Numerical values of $x_{\rm min}\equiv$REW$_{\rm min}$ that approximately represent the color-cuts adopted by Ouchi et al. (2008, see their Table~23) are given in Table~\ref{table:obs}. For the maximum REW we have assumed $x_{\rm max}\equiv$REW$_{\rm max}=300$ \AA, but note that our results do not depend precisely on this number (see \S~\ref{sec:dispar}).

The conditional PDF  for the REW [$P({\rm REW}|M_{\rm UV})$]  has been measured for LBGs. In Appendix~\ref{app:ewpdf} we show that the Ly$\alpha$ REW PDF can be well described by an exponential whose scale length depends on $M_{\rm UV}$ and $z$
\begin{equation}
P({\rm REW}|M_{\rm UV})=\mathcal{N}\exp\Big{(} \frac{-{\rm REW}}{{\rm REW}_c(M_{\rm UV})}\Big{)}.
\label{eq:pew}
\end{equation} Here $\mathcal{N}$ denotes a normalization constant, which we choose so that all drop-out galaxies have $-a_1 \leq$REW$\leq$REW$_{\rm max}$ (see Appendix~A1). In this expression the factor $a_1=20$ \AA\hs for $M_{\rm UV} < -21.5$, and $a_1=20-6(M_{\rm UV}+21.5)^2$ \AA\hs for $-21.5 \leq M_{\rm UV} \leq -19.0$, and we freeze the evolution of $a_1$ for $M_{\rm UV} > -19.0$. Furthermore, REW$_c$($M_{\rm UV})={\rm REW}_{\rm c,0}+\frac{d{\rm REW}_{\rm c}}{dM_{\rm UV}}\Delta M+\frac{d{\rm REW}_{\rm c}}{dz}\Delta z$ with $\Delta M \equiv M_{\rm UV}+21.9$, and $\Delta z=z-4.0$. The best fit values are REW$_{\rm c,0}=22\pm 3 $ \AA, and $\frac{d{\rm REW}_{\rm c}}{dM_{\rm UV}}=6 \pm 4$ \AA. Equation~(\ref{eq:pew}) ensures that the fraction of drop-out galaxies with Ly$\alpha$ in emission (i.e. REW$\gsim$ 0) depends on $M_{\rm UV}$, as is observed. We assume throughout that this fitting formula applies only in the observed range of UV magnitudes, and also freeze the evolution of REW$_c$($M_{\rm UV})$ for $M_{\rm UV} > -19.0$ (see \S~\ref{sec:dispar} for a discussion of the uncertainties this may introduce). The choice of the functional form  that approximates the data is quite arbitrary. In Appendix~\ref{app:altew} we investigate an alternative parametrization, and find that our results are not significantly affected.

\subsection{Comparison to Data}	
\label{sec:data}

We next compute `predicted' Ly$\alpha$ luminosity functions by combining Eq~\ref{eq:phi}, Eq~\ref{eq:pm1}, and  Eq~\ref{eq:pew} at $z=3.1$, $z=3.7$ and $z=5.7$. To facilitate the comparison with the data, we compute the quantity $\Psi(L_{\alpha})d\log L_{\alpha}$ which denotes the number density of LAEs in the range $\log L_{\alpha} \pm (d \log L_{\alpha})/2$. The units of  $\Psi (L_{\alpha})$ are cMpc$^{-3}$ $[\log L_{\alpha}]^{-1}$, and we have $\Psi (L_{\alpha})=\ln 10 L_{\alpha} \Phi(L_{\alpha})$. 

Each model is described by three parameters that quantify the Ly$\alpha$ REW distribution (${\rm REW}_{\rm c,0}, \frac{d REW_{\rm c}}{dM_{\rm UV}}, \frac{d REW_{\rm c}}{dz}$),  the normalization parameter $F$, and three Schechter parameters plus one cosmic variance parameter at each redshift bin. We therefore need to explore a 16-dimensional parameter space ($3+1+3\times4=16$), which we do with a Monte Carlo Markov Chain (MCMC) method. 
 We characterize each model by the parameter vector ${\bf P}=({\rm REW}_{\rm c,0}, \frac{d REW_{\rm c}}{dM_{\rm UV}}, \frac{d REW_{\rm c}}{dz}$ $,F,\alpha_{3,..,6},\phi^*_{3,...,6},M_{\rm UV,3,...,6}^*,\delta_{\rm V,3,...,6})$. Here, we have adopted a notation where  $\alpha_{3}$ denotes the value of $\alpha$ at $z=3$. We compute the posterior probability for each model as $P({\bf P}) \propto \mathcal{L}[{\bf P}]P({\bf P})$, where $\mathcal{L}[{\bf P}]=\exp[-0.5\chi^2]$ denotes the likelihood, in which $\chi^2=\sum_i^{N_{\rm data}}({\rm model}_i-{\rm data}_i)^2/\sigma^2_i$. The function $P({\bf P})$ denotes the prior PDF for these parameters \citep[e.g.][]{Cowan97}. 
\begin{figure*}
\vbox{\centerline{\epsfig{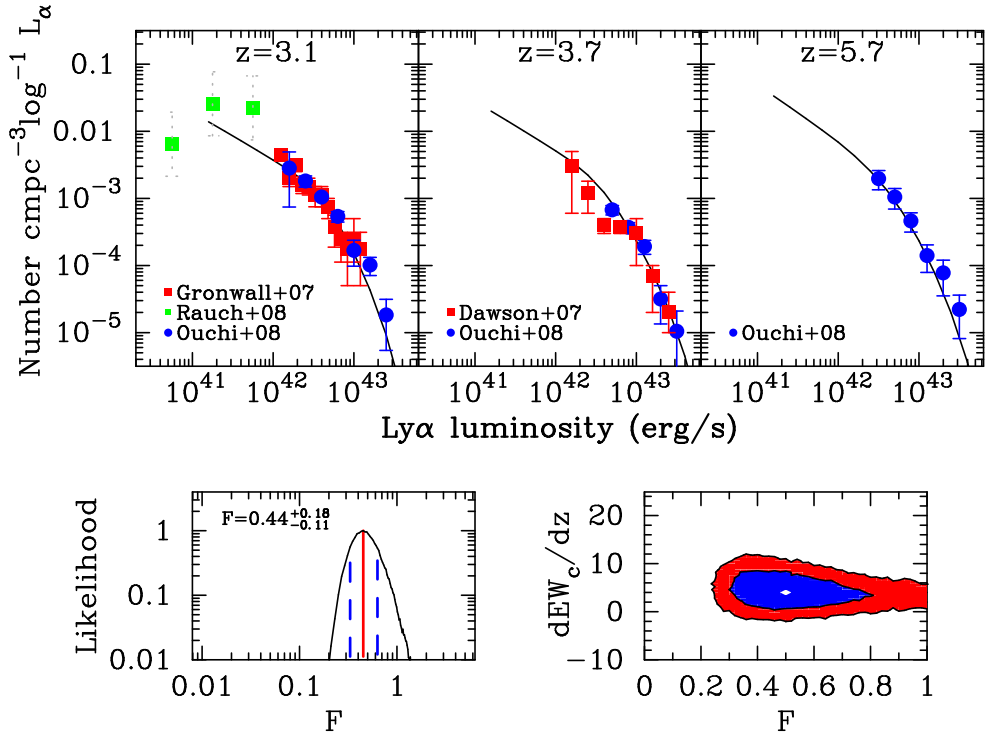}}}
\caption[]{We plot Ly$\alpha$ luminosity functions at $z=3.1$ ({\it upper left}), $z=3.7$ ({\it middle}) and $z=5.7$ ({\it upper right}). The {\it solid black lines} show our best fit model, which we calculated following the procedure described in \S~\ref{sec:dropLF} -\S~\ref{sec:data}. The data of Ouchi et al. (2008) are indicated as {\it blue circles} in {\it all panels}.  For completeness, we have also shown some other data sets, which were not included in our fits (see text). The {\it lower left panel} show marginalized posterior PDF for the parameter $F$. The {\it lower right panel} shows the 68\% and 95\% contours in the $F-\frac{d{\rm REW}_{\rm c}}{dz}$ plane, which we obtained by marginalizing over all other parameters.}
\label{fig:LF1}
\end{figure*}

We have assumed that $P({\bf P})$ is a multivariate Gaussian, i.e. $P({\bf P})=\mathcal{N}\exp\Big{[} -\frac{1}{2}({\bf P}-{\bf \mu}_{P})^T{\bf C}^{-1}({\bf P}-{\bf \mu}_{P})\Big{]}$, where $\mathcal{N}$ denotes the normalization factor. The vector ${\bf \mu}_{P}$ contains the best fit values for each of the parameters (e.g. from Table~1 we have $\mu_{\alpha,3}=-1.73$, and in \S~\ref{sec:plamuv} we find that $\mu_{\rm REW,c,0}=22$ \AA). Note that we do not assume any prior knowledge of $F$ or $\frac{d{\rm REW}}{dz}$. The covariance matrix ${\bf C}$ contains the measured uncertainties on the parameters\footnote{The covariance matrix in this case is a $16\times 16$ matrix whose entries are given by $C_{ij}=\sigma_i\sigma_j\rho_{ij}$. Here $\sigma_i$ denotes the uncertainty on parameter `$i$', and $\rho_{ij}$ denotes the correlation coefficient between parameter $i$ and $j$.  While these correlation coefficients are generally not given, the constraints on the parameters $\alpha,M_{\rm UV}^*,\phi^*$ at a given redshift {\it are} strongly correlated. We assumed throughout that $\rho_{\alpha,M^*}=\rho_{M^*,\phi^*}=\rho_{\alpha,\phi^*}=0.9$ at each redshift. We found that this decently reproduces the shape of the $68\%$ and $95\%$ likelihood contours for different Schechter parameter combinations as given by Bouwens et al. (2007).  We assume that the other parameters are not correlated, i.e. $\rho_{ij}=0$. By definition $\rho_{ii}=1$ for all `$i$'. }. Table~\ref{table:param} summarizes the assumed redshift evolution for $M^*_{\rm UV},\phi^*$ and $\alpha$ and their uncertainties. We then compute marginalized PDFs for the parameters $F$ and $\frac{d{\rm REW}}{dz}$ by marginalizing over the other 15 parameters.
The data that we use for the fits is from Ouchi et al. (2008). 

Finally, we summarize other model parameters that we do not vary as part of the MCMC calculations in Table~\ref{table:obs}.  These parameters include: ({\it i}) the approximate minimum equivalent width (REW$_{\rm min}$), and ({\it ii}) the minimum Ly$\alpha$ luminosity ($L_{\alpha}$) to which the narrowband survey of Ouchi et al. (2008) was sensitive.

\subsection{The `normalization' parameter $F$}
\label{secF}
The `normalization' parameter $F$ scales the predicted luminosity function up and down, and so can be interpreted as the ratio of the observed to predicted number density of LAEs. In this paper we focus on the value of $F$ that arrises from fitting to LAE and LBG luminosity functions, as a means of testing the validity of using a simple cut in Ly$\alpha$ luminosity and REW to represent the real observational selection criteria. In a case where the simple cuts accurately represent the true selection criteria (and hence models provide a faithful representation of the data) we therefore expect that $F=1$. However any deviation of $F$ from unity indicates that the simple cuts are not providing an adequate description of the observational selection criteria.

\section{Results}
\label{sec:results}

\subsection{The Ly$\alpha$ Luminosity Functions}
\begin{figure*}
\vbox{\centerline{\epsfig{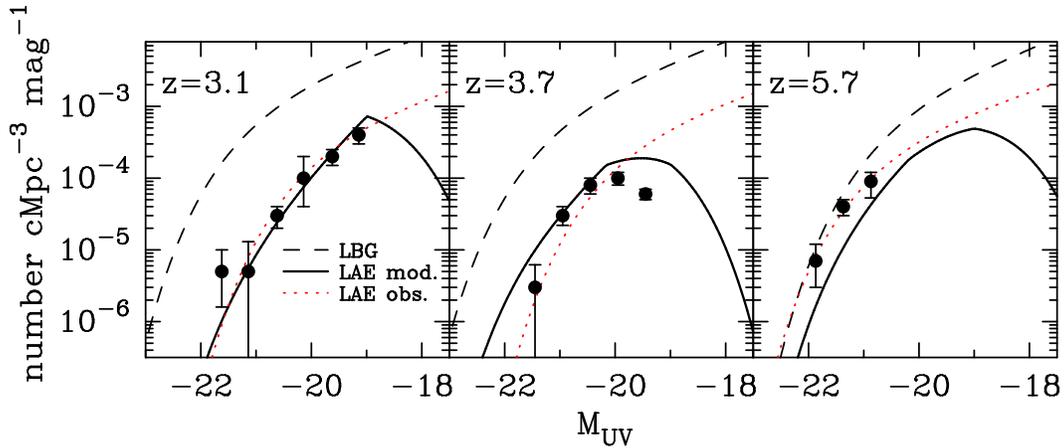}}}
\caption[]{The predicted UV luminosity function of LAEs for our best-fit model ({\it black solid lines}), compared to the data of Ouchi et al. (2008).  The agreement between our model and the data is excellent at $z=3.1$ ({\it left panel}, note that this is not a fit), and reasonable at $z=3.7$ ({\it central panels}).  Our model underpredicts the UV luminosity function at $z=5.7$, which may be related a slight overabundance of large REW (REW$\gsim 200$ \AA) systems in our model (see text). For completeness, we have also shown the best-fit drop-out galaxy luminosity function ({\it black dashed lines}), and the best fit Schechter function of the UV luminosity function of LAEs as derived by Ouchi et al. (2008, {\it red dotted lines}).}
\label{fig:UVLF}
\end{figure*}
\begin{figure*}
\vbox{\centerline{\epsfig{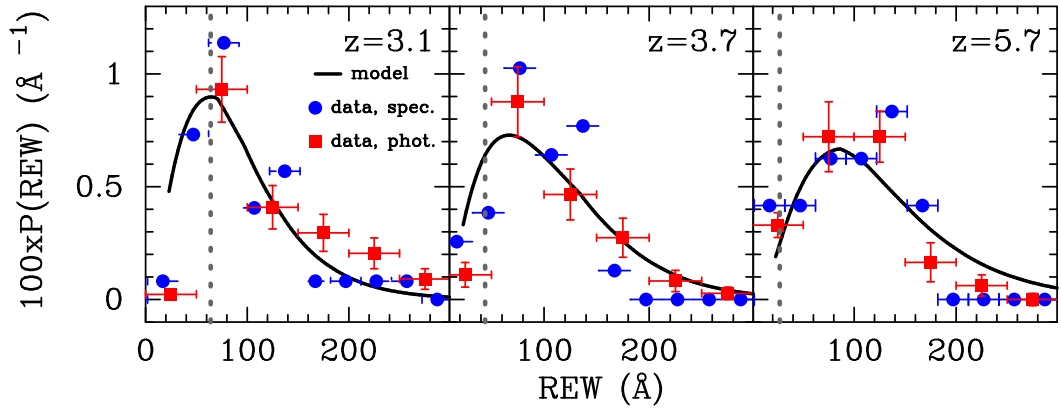}}}
\caption[]{The model equivalent width PDF ({\it black thick solid lines}), compared to the best estimate of the observed REW PDF at $z=3.1$, $z=3.7$ and $z=5.7$ (Ouchi et al. 2008),  for all their photometrically selected LAEs ({\it red filled squares}), and for spectroscopically identified LAEs ({\it blue filled circles}). The {\it vertical dotted lines} indicate the minimum REW of the Ly$\alpha$  emission line that galaxies need to have in order to be selected as LAEs by Ouchi et al. (2008).}
\label{fig:EW}
\end{figure*}

In Figure~\ref{fig:LF1} we show Ly$\alpha$ luminosity functions at $z=3.1$ ({\it upper left}), $z=3.7$ ({\it middle}) and $z=5.7$ ({\it upper right}). The {\it solid black lines} show our best fit model (see below), which we calculated following the procedure described in \S~\ref{sec:dropLF} -\S~\ref{sec:data}. The data of Ouchi et al. (2008) is indicated as {\it blue circles in all panels}. For completeness, we have also shown the $z=3$ data from Rauch et al. (2008, {\it green circles}) and Gronwall et al. (2007, {\it red circles}), and the $z=3.7$ data from Dawson et al. (2008, {\it red squares}). These other data points were not included in our fits. 

The best fit model is in excellent agreement with the data, and is described by the parameters ${\rm REW}_{\rm c,0}=23$ \AA, $\frac{d REW_{\rm c}}{dM_{\rm UV}}=7$ \AA, $\frac{d REW_{\rm c}}{dz}=6$ \AA, $F=0.53$, $\alpha_{3}=-1.65$, $\phi^*_{3}=1.9\times 10^{-3}\hs{\rm cMpc}^{-3}$,$M_{\rm UV,3}^*=-20.9$, $\delta_{\rm V,3}$=+0.04; $\alpha_{4}$=-1.70, $\phi^*_{4}=1.3\times 10^{-3}\hs{\rm cMpc}^{-3}$, $M_{\rm UV,4}^*=-20.9$, $\delta_{\rm V,4}$=-0.10, $\alpha_{6}=-1.73$, $\phi^*_{6}=1.5\times 10^{-3}\hs{\rm cMpc}^{-3}$, $M_{\rm UV,6}^*=-20.4$, $\delta_{\rm V,6}=0.01$. It is interesting -- though likely a coincidence -- that this model is in reasonable agreement with the data of Rauch et al. (2008, {\it green circles}). These authors detected 27 ultrafaint Ly$\alpha$ emitters in a long-slit, 92 hrs, observation with the ESO VLT FORS2 spectrograph, and probed an effective volume of $\sim 10^3$ cMpc$^{3}$. In such a small volume cosmic variance is significant. Following the prescription of \citet{S04} for estimating cosmic variance, we find that the uncertainty on the number of detected galaxies is $\sigma_N \sim 0.8 \langle N \rangle$. These uncertainties are denoted as {\it grey dashed lines} because this estimate is likely not accurate for the highly elongated survey volume probed by Rauch et al. (2008, see Mu{\~n}oz et al. 2010 for a discussion of cosmic variance in pencil beam surveys).

The {\it lower left panel} shows the marginalized posterior PDF for the parameter $F$. Based on available data for the drop-out galaxy population we find that we {\it overpredict} the number density of LAEs, and so need to multiply our predictions by a constant factor of $F\sim 0.43$ to match the data. Formally, $F=1$ is ruled out at $\sim 95\%$ CL. As noted previously, in a case where sharp REW and luminosity cuts among the population of  drop-out galaxies can be used to reproduce the observed luminosity functions of LAEs we expect $F=1$. The fact that the data are inconsistent with $F=1$ therefore demonstrates that the use of sharp luminosity and REW cutoffs does not adequately describe the observational selection of LAEs. This represents the primary conclusion of our paper.

In the {\it lower right panel} we plot the 68\% and 95\% contours in the $F-\frac{d{\rm REW}_{\rm c}}{dz}$ plane, which we obtained by marginalizing over all other parameters. The data clearly prefer models with $\frac{d{\rm REW}_{\rm c}}{dz}>0$, in excellent agreement with the findings of Stark et al. (2010, 2011, also see Fig~\ref{fig:dewdz}).

\subsection{The UV Luminosity Functions}

We next take our best-fit model and compare it to other observed properties of LAEs. First, we `predict' the UV luminosity function of LAEs, $\Phi_{\rm LAE}(m_{\rm UV},z)$

\begin{equation}
\Phi_{\rm LAE}(m_{\rm UV},z)=\phi(M_{\rm uv},z)\times\mathcal{F}(M_{\rm UV},z),
\end{equation} where $\phi(M_{\rm uv},z)$ denotes the LBG UV luminosity function (introduced in \S~\ref{sec:form}), and $\mathcal{F}(M_{\rm UV},z)$ denotes the fraction of drop-out galaxies that have REW$\geq {\rm REW}_{\rm min}$ and a total Ly$\alpha$ flux greater than $L_{\alpha,{\rm min}}$ (see Table~\ref{table:obs} for numerical values of $L_{\alpha,{\rm min}}$ and $ {\rm REW}_{\rm min}$). Note that this latter constraint is important. For fainter drop-out galaxies, the requirement that the Ly-$\alpha$ flux be sufficiently large can translate to a higher required minimum REW (also see Zheng et al. 2010a). The fraction $\mathcal{F}(M_{\rm UV},z)$ is then obtained by integrating over the REW-PDF.

Figure~\ref{fig:UVLF} shows the best-fit drop-out galaxy luminosity function ({\it black dashed lines}), the best fit Schechter function of the UV luminosity function of LAEs as derived by Ouchi et al. (2008, {\it red dotted lines}), and our predicted UV luminosity function of LAEs for the best-fit model shown in Figure~\ref{fig:LF1} ({\it black solid lines}). The `breaks' in the UV-luminosity functions at $M_{\rm UV}>-19$  arise because Ly$\alpha$ REWs greater than our quoted REW$_{\rm min}$ are required to render the Ly$\alpha$ flux large enough (i.e. $L_{\alpha} \geq L_{\alpha,{\rm min}}$, see Table~\ref{table:obs}) to be detected from fainter UV magnitudes. For example, in order for an object to be detected in Ly$\alpha$ from an $M_{\rm UV}=-18.5$ galaxy, we need REW$>100$ \AA, and these objects are significantly rarer. Note that the `sharpness' of the break increases towards lower redshift. This is because the scale-length of the REW-PDF increases with redshift (see Fig~1), and objects with REW$>100$ \AA\hs are rarer at lower redshift. 

The agreement between our model and the data is excellent at $z=3.1$ ({\it left panel}, note that this is not a fit), and reasonable at $z=3.7$ ({\it central panels}). In the $z=3.7$ case, the observed UV-LF suffers from incompleteness at $M_{\rm UV} > -20.0$, and so our formalism therefore overpredicts only the brightest UV point. Our best-fit $z=5.7$ model clearly underpredicts the LAE UV-LF at $z=5.7$ ({\it right panel}) at $-20< M_{\rm UV} < -21.4$. The reason for this is not clear. It is possibly related to the fact that we overproduce the number of large REW (REW$\gsim 200$ \AA) systems (see below), although this discrepancy in the REW distribution is quite small.

\subsection{The Ly$\alpha$ Equivalent Width Distribution}

We also use our formalism to predict the REW distribution (derived in Appendix~\ref{app:EWPDF}) for our best-fit model  

\begin{eqnarray}
\nonumber
P({\rm REW},z)&=&\\ 
&&\hspace{-20mm}\mathcal{N}\int_{L_{\alpha,{\rm min}}}^{L_{\alpha,{\rm max}}} P({\rm REW}|M_{\rm UV,c},z)\phi(M_{\rm UV,c},z)d\log_{10} L_{\alpha},
\end{eqnarray} where $\mathcal{N}$ denotes a normalization constant,  $\phi(M_{\rm uv},z)$ again denotes the LBG UV luminosity function (introduced in \S~\ref{sec:form}), and $P({\rm REW}|M_{\rm UV,c},z)$ denotes the Ly$\alpha$-REW PDF that is observed for LBGs. The relation between Ly$\alpha$ luminosity, UV continuum flux density, and REW uniquely determines the absolute UV magnitude $M_{\rm UV,c}$ at fixed $L_{\alpha}$ and REW. This equation states that at a given Ly$\alpha$ luminosity, the probability of observing a galaxy with Ly$\alpha$ REW is the sum of all possible $M_{\rm UV}$, weighted by their number density.

Figure~\ref{fig:EW} shows our model equivalent width PDF ({\it black thick solid lines}).  The {\it red filled squares} show the observed REW PDF for all photometrically selected LAEs (also shown as the {\it black histograms} in Fig~23 of Ouchi et al. 2008). The {\it blue dashed histograms} show the observed REW distribution of spectroscopically confirmed LAEs. At $z=3.1$ ({\it left panel}) our model underpredicts the observed number of drop-out galaxies with REW$>175$ \AA. The agreement between our model and the data at $z=3.7$ is excellent. As eluded to previously, our model slightly overpredicts the number of large REW systems (REW$>200$ \AA) at $z=5.7$, which partially explains why we underpredict the observed UV-LF of LAEs at $z=5.7$.

\section{Discussion}
\label{sec:dis}

In the previous section we showed that the observed REW distribution of LAEs can be  reproduced well out to REW=200~\AA~ using sharp cuts in REW and Ly$\alpha$ luminosity among the drop-out population. However our empirical procedure leads to LAE luminosity functions (both UV and Ly$\alpha$) that are overpredicted by a factor of $1/F\sim2.5$. As already noted, the difference of $F$ from unity indicates that simple selection cuts in luminosity and REW do not adequately represent the LAE selection. The goal of this discussion is to explore the observational biases and model assumptions that may cause $F$ to be less than $1$.

\subsection{Discussion of Observational Biases}
\label{sec:bias}

Firstly, we note that using the drop-out galaxy population to constrain the number density of star forming galaxies as a function of their UV and Ly$\alpha$ luminosity, does not miss the small fraction of  `red' star forming galaxies \citep[e.g. those with UV slopes $\beta \gsim -0.5$,][]{UVslope}, since the drop-out galaxy luminosity functions used in this paper have been corrected for this bias \citep[see][]{UVslope}. Secondly, narrowband surveys can pick up galaxies such as ULIRGs which do not make it into drop-out surveys \citep{Ni11}.  However, the ULIRG fraction among LAEs drops dramatically to $\lsim 10\%$ at $z>2.5$ \citep{Ni11}, and this is unlikely to be a significant effect. A hypothetical population of  LAEs whose Ly$\alpha$ emission is powered predominantly by gravitational heating (as in e.g. Birnboim \& Dekel 2003, Dijkstra 2009, Dayal et al. 2010), would also contribute to LAE samples. However these known potential observational biases would enhance the true number density of LAEs, and thus lead to a value of $F>1$. 

Ouchi et al. (2008) discuss possible explanations why $F >1$ (see \S~\ref{sec:other} for a more detailed comparison with Ouchi et al. 2008). Some of these possibilities could also result in $F<1$. These include: ({\it i}) Systematic uncertainties in drop-out galaxy luminosity functions, based the observed scatter in the $z=6$ drop-out LFs obtained by different groups. ({\it ii}) The fact that drop-out LFs at a particular redshift are measured over a much broader redshift interval ($\Delta z \sim 1$) than is probed by narrowband surveys. These differences may introduce extra uncertainties when comparing LAE and drop-out galaxy populations. 

We have investigated whether our results are dominated by observations at a particular redshift, and repeated our analysis based on the LAE luminosity function in individual redshift bins (while fixing $\frac{{\rm d REW}_{\rm crit}}{dz}=6$\hs\AA). We find that $F=0.34^{+0.46}_{-0.17}$ at $z=3.1$,  $F=0.44^{+0.48}_{-0.21}$ at $z=3.7$ and $F=0.52^{+0.48}_{-0.23}$ at $z=5.7$, where the uncertainty on $F$ is dominated by cosmic variance. Thus, our result is not dominated by any redshift bin.  The systematic effect that causes $F<1$ appears to operate at all redshifts.

 \subsection{Discussion of Model Assumptions}
\label{sec:dispar}

In this section we discuss the effects of different assumptions on our results, with particular focus on the conclusion that $F<1$. 

{\bf Assumption 1.}  In our fiducial model we assume that the fitting formula for REW (Eq~\ref{eq:pew}) applies only in the observed range of UV magnitudes, and freeze its evolution for $M_{\rm UV} > -19.0$. On the other hand, Stark et al. (2010) find that the observed evolution in the REW-PDF continues down to $M_{\rm UV} = -18.5$ (see the {\it left panel} of their Fig~13), albeit with large uncertainties. Had we extrapolated our fitting function down to fainter UV magnitudes, then we would have allowed more UV-faint, large REW galaxies into our sample. However, if we include these UV-faint sources, then we would push our constraints on $F$ to lower values, which would rule out $F=1$ at greater significance. Conversely, the uncertainties on the REW-PDF are large at $M_{\rm UV} \geq -19.0$. If we had frozen the evolution of the REW-PDF at $M_{\rm UV} \geq -19.25$, then we would have found a larger value of $F$, especially when combined with a truncation of the REW-PDF at REW$\geq 150$ \AA\hs (see below).

{\bf Assumption 2.} In our fiducial model we chose $x_{\rm max}=$REW$_{\rm max}=300$ \AA ~to be the maximum possible Ly$\alpha$ REW. This value is close to the largest REW in the sample of Ouchi et al. (2008).  Theoretically, the Ly$\alpha$ REW can reach $\sim 1500-3000$ \AA\hs for metal free galaxies forming stars with a top-heavy IMF (Schaerer  2003, Johnson et al. 2009, Raiter et al. 2010), or cooling clouds (Dijkstra, 2009), and could be boosted to even larger values if dust preferentially suppresses the UV continuum \citep[][]{Neufeld91,Hansen06}. Adopting larger values for REW$_{\rm max}$ would boost the overall Ly$\alpha$ emissivity of star forming galaxies, which would again reduce our best-fit values for $F$.

We note that there is limited data to support the assumed exponential form of the REW-PDF at REW$\gsim 150$ \AA. Indeed,  among the $\sim 800$ LBGs in the sample of Shapley et al. (2003), only 4 (1) have REW$\geq 150$  (175) \AA, so that the observed REW-PDF among LBGs is very uncertain at these values.  If we truncate the REW-PDF at REW$_{\rm max}=150$ \AA, then we find $F=0.64^{+0.20}_{-0.12}$, and $F=1$ is only ruled out at $\sim 89\%$ CL. If we further combine this with `freezing' the evolution of the REW-PDF at $M_{\rm UV} \geq -19.25$, then we find $F=0.70^{+0.23}_{-0.14}$, and $F=1$ is only ruled out at $\sim 77\%$ CL. While these modified assumptions do not fully resolve the issue, they do illustrate that our finding of $F\ll1 $ depends on the uncertain REW-PDF at REW$\gsim 150$ \AA\hs and faint $M_{\rm UV}$. We note that if we assume REW$_{\rm max}=150$ \AA, then our model does not produce LAEs with REW$\geq 150 \AA$, although these objects {\it are} observed. However, considering just LAEs that have been confirmed spectroscopically, this appears to be a significant problem only at $z=5.7$, where the uncertainties on measured REWs are large (see Table~2 of Ouchi et al. 2008).

{\bf Assumption 3.} In our fiducial model we assumed that $\beta=-1.7$. \citet{Stark10} found that $\beta=-1.6$ for drop out galaxies with REW$<50$ \AA, and $\beta=-2.0$ when REW$\geq 50$ \AA\hs for $-21.5 <M_{\rm UV}<-20.5$. For fainter galaxies with REW$<50$ \AA, $\beta$ approaches $-2.0$. Our choice for $\beta$ may be slightly too high \citep[also see][]{UVslope,Bouwens10b}. However, if we decrease $\beta$ then we {\it increase} the total Ly$\alpha$ flux for a fixed REW and $M_{\rm UV}$, and thus the overall Ly$\alpha$ emissivity of the drop-out galaxy population. Decreasing $\beta$ would therefore again lower our overall best-fit value for $F$.

\subsection{Comparison to Previous Work}
\label{sec:other}
Our study bares similarities to that of Malhotra \& Rhoads (2002). These authors also compared observations with `predicted' LAE number counts and REW distributions, which they obtained by combining existing constraints on the faint-end of the $z\sim4$ drop-out galaxy luminosity function with a theoretical model for the Ly$\alpha$ REW-distribution. However, modeling the observed equivalent width of the Ly$\alpha$ emission line is a complicated task which depends on detailed radiative transfer of Ly$\alpha$ photons through both ISM and IGM. Our calculations completely circumvent this complication by utilizing empirical distributions, which represents an important improvement. In agreement with our findings, the study of \citet{MR02} also overpredicted the number density of LAEs. Their offset was by an even larger factor of $\sim 6-12$, implying that only $\sim 7-15\%$ of all LBGs need to be LAEs. However our formalism already includes the observation that only a fraction of the drop-out population has strong enough Ly$\alpha$ emission to qualify as a LAE. Thus the concept of separate galaxy populations cannot be invoked to explain the offset ($F\neq1$). \\

Ouchi et al. (2008) found that the UV-LF of the LAEs at $z=5.7$ lies remarkably close to the UV-LF of drop-out galaxies at $z=6$, which suggests that $\sim 50-100\%$ of the drop-out galaxies at z=6 would qualify as LAEs (also see Shimasaku et al. 2006). Phrased alternatively, the UV-LF of LAEs implies that $\sim 50-100\%$ of the  drop-out galaxies have a Ly$\alpha$ emission line whose REW exceeds REW$_{\rm crit}\approx 20$ \AA. This conflicts with spectroscopic observations of LBGs at that redshift which suggest the observed number is closer to $\sim 30\%$ \citep{Shapley03,Stanway07,DH07}.

The $z=5.7$ UV-LF of Ouchi et al. (2008) therefore implies that $F\approx [50-100\%]/ 30\% =1.6-3.3$, i.e. they find {\it more} LAEs than expected from drop-out galaxy populations.  At first glance this conflicts with our finding that $F<1$.  However, our constraint on $F$ was derived by considering the Ly$\alpha$ LFs at $z=3.1$,  $z=3.7$ and $z=5.7$. We showed in \S~\ref{sec:bias} that $F=1$ was only excluded at the $1-\sigma$ level if we had only considered the $z=5.7$ Ly$\alpha$ LF. Importantly, the $z=5.7$ UV-LF is the {\it only} luminosity function of the six LFs that we modelled for which $F=1$ would have given a good fit (Figure~\ref{fig:UVLF} shows that our best-fit model undershoots the UV-LF by a factor of $\sim 2$, despite the fact that our model reproduces the Ly$\alpha$ LF and REW distributions at this redshift as well as it does at other redshifts). 

The $z=5.7$ UV-LF alone therefore appears to be consistent with direct spectroscopic observations of LBGs. This still seems at odds with Ouchi et al. (2008) who were concerned with explaining why $F$ significantly exceeded unity. The discussion in Ouchi et al. (2008) used observational constraints on the REW-PDF of the drop-out galaxy population that were available at that time. However recent data which contains larger samples of galaxies  shows that the fraction of drop-out galaxies with REW$>20$ \AA\hs increases dramatically at $M_{\rm UV} > $-20.0, which is relevant when comparing to LAEs. If one accounts for this increase, then the discrepancy noted by Ouchi et al.~(2008) becomes less serious. Indeed, in the recent compilation by Ono et al. (2011), the fraction of faint drop-out galaxies for which REW$\geq 25$ \AA\hs is $\sim 55 \pm 15\%$. It therefore seems likely that both this work and that of Ouchi et al. (2008) would conclude that the UV-LF of $z=5.7$ LAEs is consistent with direct spectroscopic observations of LBG. In contrast with previous studies, our constraints are derived from a model that uses more available data on the Ly$\alpha$ REW-PDF observed in drop-out galaxies, considers the Ly$\alpha$ luminosity functions, and studies different redshifts.

\section{Extrapolating to Redshifts $z\sim 7-9$}
\label{sec:z9}

\begin{figure}
\vbox{\centerline{\epsfig{file=fig4.eps,angle=270,width=8.0truecm}}}
\caption[]{The fraction of drop-out galaxies with Ly$\alpha$ emission lines with REW$>x$ as a function of redshift.  {\it Blue filled squares} show the data from Stark et al. (2010) for galaxies with $-20.5 < M_{\rm UV} < -19.5$ and $x=75$ \AA, while {\it red filled circles} show the data from \citet{Stark10,P11,S11,P11} compiled by Ono et al. (2011) for $-20.25 < M_{\rm UV} < -18.75$ and $x=25$ \AA. The {\it solid lines} shows our best fit model (this model was fitted to the Ly$\alpha$ luminosity functions, not this data), which provides and excellent fit to the data for our purposes (see text). The {\it dotted lines} show our best-fit model extrapolated to $z \geq 6$.  The observed fraction at $z=7$ falls well below this extrapolation.}
\label{fig:dewdz}
\end{figure}

\begin{figure*}
\vbox{\centerline{\epsfig{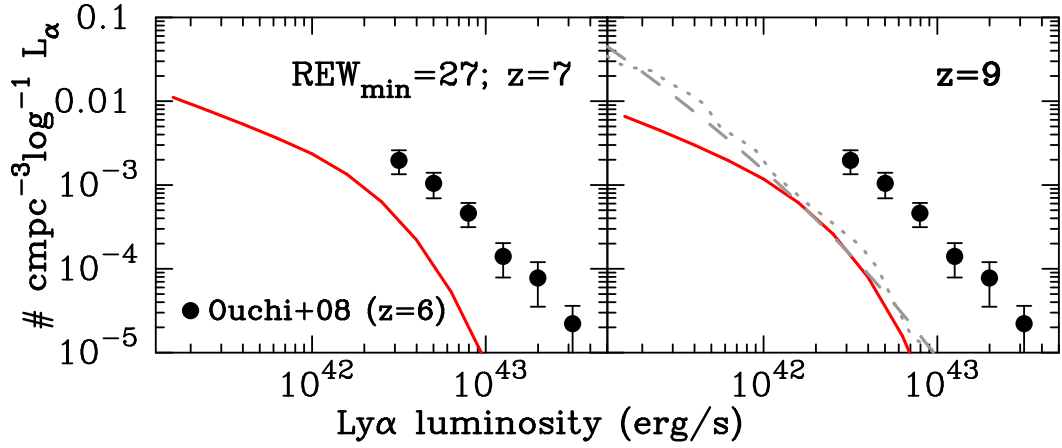}}}
\caption[]{{\it Left Panel:} Predicted number density of LAEs (in cMpc$^{-3}$ log$^{-1}L_{\alpha}$) at $z=7$ ({\it solid red line}). {\it Right panel:} Same as the {\it left panel}, but at $z=9$. Within the range of luminosities that will likely be probed by future surveys, $L_{\alpha}\sim 10^{41}-10^{43}$ erg s$^{-1}$, the agreement between previous model predictions (here {\it dashed line} is the GALFORM model from Nilsson et al. 2007, and the {\it dotted line} is the model of Dijkstra et al. 2007a) and those inferred from the z=9 LBG population is quite good (see text).}
\label{fig:highz}
\end{figure*}

There are existing constraints on the LBG luminosity functions at $z>6$. It is therefore interesting to take our best-fit model and predict the expected number density of LAEs at redshifts beyond those currently observed. Before doing so, we first point out that extrapolating the REW-PDF of our best-fit model to $z>6$ leads to inconsistencies with some available data. 

The observed fraction of drop-out galaxies having Ly$\alpha$ emission lines of REW$\geq x$ has been measured  as a function of redshift out to $z\sim7$ for galaxies with $-20.5 < M_{\rm UV} < -19.5$ for $x=75$ \AA\hs\citep{Stark10}, and for galaxies with $-20.25 < M_{\rm UV} < -18.75$ for $x=25$ \AA\hs\citep[][also see Vanzella et al. 2011]{P11,S11,Ono11}. In Figure~\ref{fig:dewdz}, we compare the data from Stark et al. (2010, {\it blue filled squares}), and the data compiled by Ono et al. (2011, {\it red filled circles}) with our best-fit model of \S~\ref{sec:results} ({\it solid lines}). Our best fit-model slightly overpredicts (by a factor of $\sim 1.2$) the fraction of drop-out galaxies for which REW$\geq 25$ \AA\hs at $z\leq 4$. However, at these redshifts REW$_{\rm min}\geq 44$ \AA, and our results depend weakly on this discrepancy. For our purposes, the agreement between our model and the data is excellent out to $z\sim6$. We stress that this best-fit model was fitted to the Ly$\alpha$ luminosity functions, and not these particular data points. This indicates that our model reproduces the observed redshift {\it evolution} of LAEs well, despite the fact that the overall predicted number density is off by a factor of $F$. However observations indicate that a sudden drop occurs in the `LAE fraction' at $z>6$, and the linear extrapolation of our model does not reproduce this evolution. 

To remain consistent with the observed drop in the `LAE fraction', we replace the term REW$_{\rm c,0}+\frac{d{\rm REW}_{\rm c}}{dz}\Delta z$ (yielding REW$_{\rm c,7}=43$ \AA\hs), in our best fit model with REW$_{\rm c,7}=8$ \AA\hs. This ensures that we reproduce the observed drop in the LAE fraction. Otherwise, we take the Schechter function parameters given in Table~\ref{table:param}, and the model parameters from our best-fit model. In the {\it left panel} of Figure~\ref{fig:highz} the {\it red solid line} shows the predicted number density of LAEs (in cMpc$^{-3}$ log$^{-1}L_{\alpha}$) at $z=7$, where we further assumed REW$_{\rm min}=27$ \AA  (to facilitate the comparison with the $z=5.7$ data by Ouchi et al. 2008). Our model predicts the cumulative number density of LAEs brighter than $L_{\alpha}=10^{43}$ erg s$^{-1}$ to be $n(L_{\alpha}>10^{43}\hs{\rm erg/s})\sim 10^{-6}$ cMpc$^{-3}$, which is below the observational constraints by Ota et al. (2010). However the observational constraints are uncertain given that their observed cumulative luminosity function is derived from only three objects. Our predicted luminosity functions are in turn affected by uncertainties in our model parameter vector ${\bf P}$. For example, taking REW$_{\rm c,7}=15$ \AA--which gives a Ly$\alpha$ fraction among $z\sim7$ drop-out galaxies of $35\%$-- results in $n(L_{\alpha}>10^{43}\hs{\rm erg/s})\sim 5 \times 10^{-6}$ cMpc$^{-3}$, which is within $1-\sigma$ of the best-fit value derived by Ota et al. (2010). Note that similar uncertainties will apply to any theoretical model, as these are likely to be --just like our parameter vector ${\bf P}$--calibrated by lower-redshift data.

The {\it right panel} of Figure~\ref{fig:highz} shows our predictions for redshift $z=9$. We assumed $M^*_{\rm UV}=-19.1$, which we obtained by extrapolating the observed redshift evolution at lower redshift ($z\leq 6$) $M^*_{\rm UV}=-21.02+0.36(z-3.8)$ \citep{Bouwens08}, and we assumed that REW$_{\rm c,9}=$REW$_{\rm c,7}$. This extrapolated value is consistent with the derived lower limit  $M^*_{\rm UV}\gsim -19.6$ derived by \citet{Bouwens08} based on the absence in J-dropouts in the Hubble Ultra Deep Field (see Oesch et al. 2011 for constraints on the $z=10$ drop-out galaxy luminosity function using deeper and wider WFC3-IR data). For completeness, we have overplotted some theoretical predictions for the number density of LAEs at $z=9$. The {\it dotted line} shows the predicted number density of LAEs at $z=8.8$ by \citet{Ni07} who used the semi-analytic model GALFORM \citep{Cole00}. The {\it dashed line} was obtained from a simpler model in which some fraction $f_*$ of all baryons is converted into stars over a timescale $\epsilon_{\rm DC}t_{\rm hub}$ \citep[][]{LF}. These simpler models easily reproduce the observed number density of LAEs at $z \geq 5.7$, and have been used frequently in the recent literature. The theoretical predictions also agree well with each other, which is probably because the models are calibrated by the same lower redshift data. The theoretically predicted luminosity functions decrease more steeply than those obtained from the LBG population. However, given the present-day uncertainties on the $z=9$ LBG luminosity function, it is not clear how significant this difference is. For example, a steeper faint end slope of the $z=9$ luminosity function would would reduce the discrepancy. Observations indicate that this faint end slope may indeed become steeper at higher redshifts \citep[][also see Jaacks et al. 2011]{slope}.

\section{Conclusions}
\label{sec:conc}

In this paper we have investigated the implications of the assumption often used in theoretical modelling that the LAEs are selected using cuts in REW and luminosity. These cuts are only coarse approximations to the detailed criteria that are employed in observational studies. To quantify the importance of the approximation, we investigated whether we can reproduce LAE data if we define LAEs using a (redshift-dependent) Ly$\alpha$ luminosity and EW threshold, and a empirical model for the number density of star forming galaxies as a function of their UV and Ly$\alpha$ luminosity. We constructed this model by combining observed luminosity functions of drop-out galaxies, with the observed rest-frame equivalent width (REW) probability distribution function (PDF) of drop-out galaxies at $z=3-7$, and `predicted'  the resulting Ly$\alpha$ luminosity function at $z=3.1$, $z=3.7$ and $z=5.7$. We also use our formalism to predict the UV LF of Ly-$\alpha$ emitters, and the EW observed distribution of Ly-$\alpha$ emitters. 

As part of our analysis we demonstrate that the observed REW-PDF of $z=3$ LBGs is well described by an exponential function at REW$\geq 0$, i.e. $P({\rm REW}) \propto \exp[-{\rm REW}/{\rm REW}_{\rm c}(M_{\rm UV})]$. The scale length depends on absolute UV-magnitude, and we use recent data from Stark et al. (2010, 2011) to constrain this dependence. Using this empirical distribution of REW in LBGs, we find that we can reproduce the observed REW distribution for LAEs quite well out to REW$\sim 200$ \AA. However, in order to reproduce the LAE luminosity functions, we find that we must re-scale the predicted luminosity functions downward by a factor of $F=0.43^{+0.14}_{-0.07}$. Formally, a value of $F=1$, which is expected if the simple REW and luminosity cuts are accurate, is ruled out at $\sim 95\%$.  We found that this discrepancy can be reduced if we truncate the Ly$\alpha$ REW-PDF at REW$ \geq 150$ \AA, and `freeze' its evolution at $M_{\rm UV} \gsim -19.25$. The sample of Shapley et al (2003) only contains 4 drop-out galaxies (out of 797) with REW$ \geq 150$ \AA, and observationally the REW-PDF is constrained very poorly at these large values for REW. For this truncated REW-PDF, which stops evolving at $M_{\rm UV} \gsim -19.25$, we find $F=0.70^{+0.23}_{-0.15}$, and $F=1$ is ruled out only at $\sim 77\%$.

On the other hand, we found that the overall redshift evolution of the LAEs was reproduced very well by our empirical model. Encouraged by this result, we combine the best-fit model in our formalism with recent observed constraints of the `LAE fraction' and predict number densities of LAEs at $z=7$ and $z=9$. Current measurements of the LBG luminosity functions translate to Ly$\alpha$ luminosity functions that are consistent with observed number counts of LAEs at $z=7$, but which are still very uncertain. Nevertheless the UV LF of LBGs can be used to provide empirical guidance for future surveys aiming to discover Ly-$\alpha$ galaxies at the highest redshifts. 

We  conclude that modeling LAEs with simple REW and luminosity cuts can lead to (significant) changes to the predicted number density of LAEs. Theorists will therefore need to make more careful account of observational selection in order to produce reliable models of the observed population. This will have to include application of the proper filter transmission curves to generate mock data from the models, {\it and} then correct for these filter transmission curves following the same procedures as followed by the observers whose data one tries to reproduce. We are attempting to address these issues in more detail in on-going work. Furthermore, our work has shown that to in order to address this issue in more detail, it will be important to reduce the observational uncertainties associated with the REW-PDF of drop-out galaxies at large REW and/or faint $M_{\rm UV}$.
  
{\bf Acknowledgements} Part of this research was supported by Harvard University funds. We thank Masami Ouchi and Alice Shapley for sharing their data in tabulated form. We thank Kim Nilsson for providing us with the model luminosity functions presented in Nilsson et al. (2007). MD thanks Roderick Overzier for bringing the work of Stark et al. (2010) to our attention, and Eduard Westra, Bram Venemans and Dan Stark for helpful and stimulating discussions. We thank Masami Ouchi for valuable feedback on an earlier version of this paper, and an anonymous referee for helpful, constructive feedback.

\appendix

\section{Accuracy of Fitting Formula for observed EW-PDF}
\subsection{Fiducial REW-PDF Parameterization}

\label{app:ewpdf}
\begin{figure}
\vbox{\centerline{\epsfig{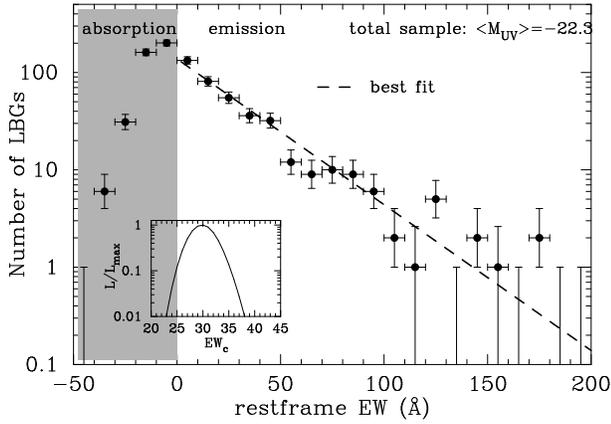}}}
\caption[]{The observed Ly$\alpha$ rest frame equivalent width (REW) distribution of $\sim 800$ $z\sim3$ LBGs of the sample of Shapley et al. (2003) is shown as the {\it black histogram}. This sample had a median $M_{\rm UV}=-22.3$. This figure shows that the observed REW-PDF can be described well by an exponential function.}
\label{fig:appewpdf0}
\end{figure}

\begin{figure}
\vbox{\centerline{\epsfig{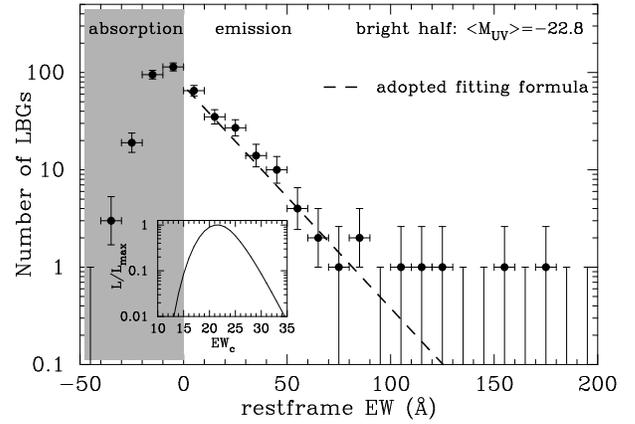}}}
\caption[]{The observed Ly$\alpha$ rest frame equivalent width (REW) distribution of the brightest 399 $z=3$ LBGs of the sample of Shapley et al. (2003) is shown as the {\it black histogram}. This sample had a median $M_{\rm UV}=-22.8$, and is labeled the 'bright' sample. The {\it dashed line} shows our adopted fitting formula (Eq~\ref{eq:pew}). Our fitting formula provides a decent fit to the data.}
\label{fig:appewpdf1}
\end{figure}
\begin{figure}
\vbox{\centerline{\epsfig{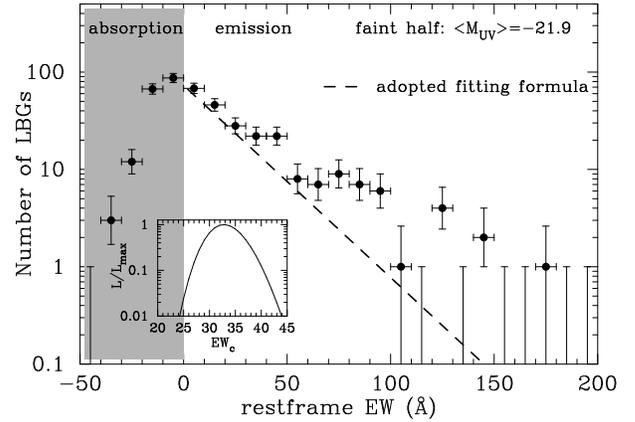}}}
\caption[]{Same as Figure~\ref{fig:appewpdf1}, but for the faintest 398 LBGs. Our fitting function now underpredicts the number of large EW systems significantly. This is because our fitting formula adopts REW$_{\rm c}=22$ \AA, while the best fit REW$_{\rm c}=33$ \AA. Our fitting formula {\it underpredicts} the number of large EW systems. Applying a correction for this would lower our required values for $F$ (see text), strengthening the conclusions of this work.}
\label{fig:appewpdf2}
\end{figure}

\begin{figure}
\vbox{\centerline{\epsfig{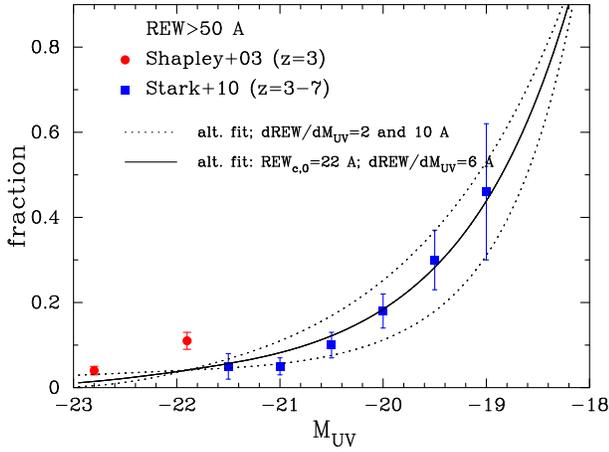}}}
\caption[]{This Figure shows the fraction of drop-out galaxies with a Ly$\alpha$ REW $> 50$ \AA, as a function of $M_{\rm UV}$. The {\it red circles} represent the $z=3$ data of Shapley et al. (2003, also shown in Fig~\ref{fig:appewpdf1} and Fig~\ref{fig:appewpdf2}). The {\it blue squares} represent data from Stark et al. (2010) from a Keck spectroscopic survey of $z=3-7$ drop-out galaxies. The {\it black solid line} shows our fitting function. The {\it black dotted lines} show our fitting function when we  increase or decrease the parameter $\frac{dREW}{dM}$ by 4 \AA. Our fitting function is clearly consistent with the data (see text).}
\label{fig:stark}
\end{figure}

In our paper we approximate the observed Ly$\alpha$ REW distribution using the functional form given by Eq~\ref{eq:pew}.  This functional form is motivated by several factors.

\begin{itemize}

 \item  The observed REW-PDF of $z=3$ LBGs is well described by an exponential function. This is illustrated in Fig~\ref{fig:appewpdf0}, where the {\it histogram} shows the observed number of LBGs as a function of Ly$\alpha$ REW (Shapley et al. 2003). This sample had a median $M_{\rm UV}=-22.3$.  The {\it inset} of this Figure shows that the best-fit scale-length associated with this exponential distribution is REW$_{\rm c}\sim 29$ \AA, for which the function is overplotted as the {\it dotted line}.

\item  The scalelength REW$_{\rm c}$ is observed to be a function of absolute UV magnitude, $M_{\rm UV}$ (Shapley et al. 2003). This is illustrated in Figures~\ref{fig:appewpdf1} and \ref{fig:appewpdf2}, where show the REW-PDF for subsamples of the brightest $\sim 400$ LBGs (Fig~\ref{fig:appewpdf1}), and of the faintest 400 LBGs (Fig~\ref{fig:appewpdf2}). The {\it thick dashed lines} show our adopted fitting formula. The figure shows that our fitting formula provides a good fit to the 'bright' sample, but significantly underpredicts the number of large EW systems in the `faint' sample (Fig~\ref{fig:appewpdf2}). This is because our fitting formula adopts REW$_{\rm c}=22$ \AA, while the best fit to this subsample is REW$_{\rm c}=33$ \AA\hs (see the {\it inset}). On the other hand this choice is required in order to be consistent with data at fainter UV magnitudes (see below). We stress that our fitting formula {\it underpredicts} the number of large EW systems at faint UV magnitudes. Applying the correction would lower our required values for $F$ (see text), strengthening the conclusions of this work.

 \item We choose the normalization\footnote{This factor is given by $\mathcal{N}=\Big{[}\exp \Big{(}\frac{a_1}{{\rm REW}_c(M_{\rm UV})} \Big{)}-\exp \Big{(}\frac{-{\rm REW}_{\rm max}}{{\rm REW}_c(M_{\rm UV})} \Big{)}\Big{]}/{\rm REW}_c(M_{\rm UV})$.} factor such that all drop-out galaxies have $-a_1 \leq$REW$\leq$REW$_{\rm max}$. While there are drop-out galaxies with smaller REW, this choice automatically results in a fraction of drop-out galaxies with Ly$\alpha$ {\it in emission} (i.e. REW $>$ 0) that increases with $M_{\rm UV}$, as observed in the sample of Shapley et al (2003). This choice for the normalization constraint allows us to describe the $M_{\rm UV}$--dependence of the observed shape and normalization of the Ly$\alpha$ REW PDF at REW$>0$ with one single parameter, namely REW$_{\rm c}$ (we keep $a_1$ constant within this range). However, as we show next this single parameter description breaks down at fainter ($M_{\rm UV} \gsim -21.5$) magnitudes.

\item Stark et al. (2010) found that the REW-PDF is sensitive to $M_{\rm UV}$. This dependance is relatively weak in the range $-22.0< M_{\rm UV} < -20.5$, but strong from $-20.5 < M_{\rm UV} < -18.5$.  This is shown in Fig~\ref{fig:stark}, where we compare the observed fraction of drop-out galaxies with REW$> 50$ \AA\hs as a function of $M_{\rm UV}$ to our fit. The {\it red squares} show the data presented by Shapley et al. (2003), while the {\it blue circles} show the data presented by Stark et al. (2010). The {\it solid curve} represents our adopted fitting function. We match the rapid evolution at $M_{\rm UV} >-20.5$ by decreasing $a_1$ as described in the paper. The {\it dotted curves} show the model when we increase or decrease the parameter $\frac{dREW}{dM}$ by 4 \AA. Note that the data point at $M_{\rm UV}=-21.9$ lies a few $\sigma$ above our fitting function, because this data point lies significantly above the data at $-21.5 \leq M_{\rm UV} \leq -21.0$. Forcing better agreement with this data point  therefore automatically results in worse agreement with the data at fainter UV magnitudes. This is why our best-fit model does not provide the best fit to the data shown in Figure~\ref{fig:appewpdf2}.

\end{itemize}
 
\subsection{Alternative Parametrization of the EW-PDF}
\label{app:altew}

\begin{figure}
\vbox{\centerline{\epsfig{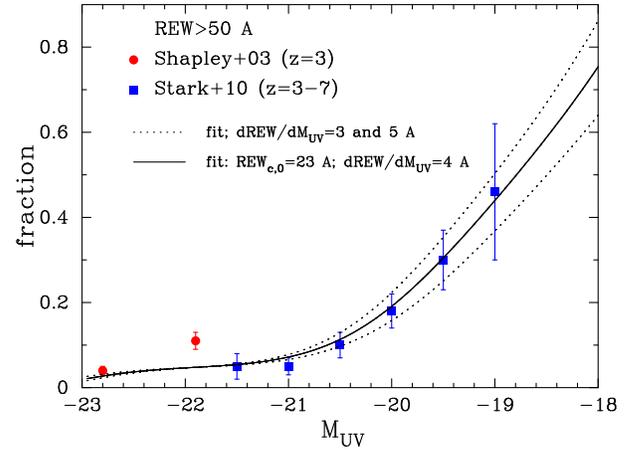}}}
\caption[]{Same as Figure~\ref{fig:stark}, but for our alternative parametrization of the REW-PDF (see text).}
\label{fig:starkalt}
\end{figure}

To make sure our results do not depend on our chosen functional form, we also study an alternative parametrization of the Ly$\alpha$ REW PDF. In this model, parametrize the observed shape and normalization of the Ly$\alpha$ REW PDF at REW$>0$ at all $M_{\rm UV}$ with a single parameter (REW$_{\rm c}$). To capture the observed evolution of the REW-PDF with $M_{\rm UV}$ (weak in the range $-22.0< M_{\rm UV} < -20.5$, and strong from $-20.5 < M_{\rm UV} < -18.5$), we add a cubic term $(\Delta M)^3$ into the expression for REW$_{\rm c}(M_{\rm UV})$. Specifically, we keep the parameter $a_1=20$ \AA\hs fixed at all $M_{\rm UV}$, and REW$_c$($M_{\rm UV})={\rm REW}_{\rm c,0}+b(\Delta M+[\Delta M]^3)+\frac{d{\rm REW}_{\rm c}}{dz}\Delta z$. The data is well described by REW$_{\rm c,0}=23 \pm 2$ \AA\hs and $b=4.4 \pm 1$\AA\hs (see Figure~\ref{fig:starkalt}).

\begin{figure*}
\vbox{\centerline{\epsfig{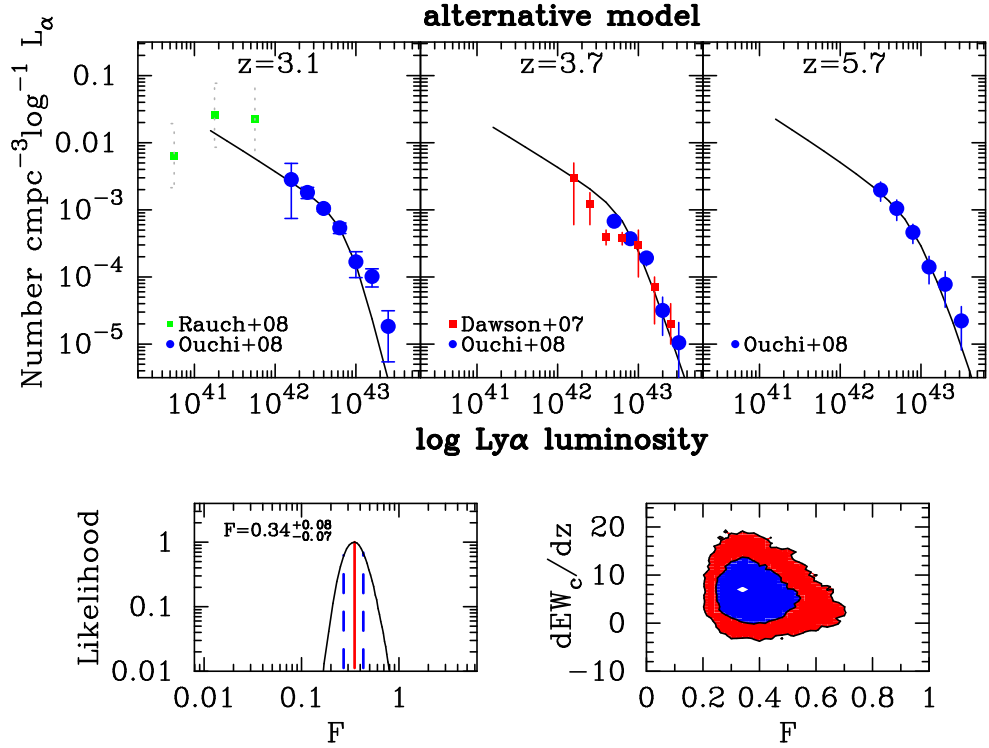}}}
\caption[]{Same as Figure~\ref{fig:LF1}, but for our alternative parametrization of the REW-PDF. The most likely value for $F$ decreased further from $F=0.43$ to $F=0.34$, and $F=1$ is ruled out at $\gsim 99\%$ CL.}
\label{fig:LF2}
\end{figure*}

We have repeated our analysis by fitting the observed Ly$\alpha$ luminosity function for this alternative model. In our standard model (see \S~\ref{sec:dispar}) we keep the evolution of the REW-PDF constant at $M_{\rm UV} \ge -19.0$. The results of this analysis are presented in Figures~\ref{fig:LF2}, which shows that our constraints on $F$ and $\frac{d{\rm REW}_{\rm c}}{dz}$ are very similar to the results already  presented in the paper. We also found  good fits for the UV luminosity functions. This gives us confidence that our results are not sensitive to the precise choice of the functional form that was used to model the REW-PDF.

\section{Calculation of EW-PDF for Ly$\alpha$ Selected Galaxies}
\label{app:EWPDF}

The EW-PDF of galaxies as a function of Ly$\alpha$ luminosity, is 

\begin{equation}
P({\rm REW}|L_{\alpha})=\int_{0}^{-\infty}dM_{\rm UV} P({\rm REW}|L_{\alpha},M_{\rm UV})P(M_{\rm UV}|L_{\alpha}).
\end{equation} 
Bayes theorem states that

\begin{equation}
P(M_{\rm UV}|L_{\alpha})=
\frac{P(M_{\rm UV},L_{\alpha})}{P(L_{\alpha})}=\frac{P(L_{\alpha}|M_{\rm UV})P(M_{\rm UV})}{P(L_{\alpha})},
\end{equation}  and we may write
\begin{eqnarray}
\nonumber
P({\rm REW}|L_{\alpha})&=&\\ 
&&\hspace{-25mm}\int_{0}^{-\infty}dM_{\rm UV}\hs P({\rm REW}|L_{\alpha},M_{\rm UV})P(L_{\alpha}|M_{\rm UV})\frac{P(M_{\rm UV})}{P(L_{\alpha})}.
\end{eqnarray}
We also know that $P(L_{\alpha}|M_{\rm UV})=P({\rm REW}|M_{\rm UV})\frac{\partial {\rm REW}}{\partial L_{\alpha}}=P({\rm REW}|M_{\rm UV})\frac{{\rm REW}}{L_{\alpha}}$ (see \S~\ref{sec:plamuv}). For a particular combination of $L_{\alpha}$ and $M_{\rm UV}$, the REW is fixed, implying that $P({\rm REW}|L_{\alpha},M_{\rm UV})=\delta_{\rm D}(g(M_{\rm UV})$, where $\delta_D(x)$ denotes the Dirac delta function and $g(M_{\rm UV})\equiv {\rm REW}-\frac{ L_{\alpha}}{C\times {\rm REW}\times L_{{\rm UV},\nu}}$. We can therefore write
\begin{eqnarray}
\nonumber
P({\rm REW}|L_{\alpha})&=& \int_{0}^{-\infty}dM_{\rm UV}\hs \delta_{\rm D}(g(M_{\rm UV}))P(M_{\rm UV})  \\ 
 \nonumber
&&\times P({\rm REW}|M_{\rm UV}) \frac{{\rm REW}}{L_{\alpha}P(L_{\alpha})}\\
 \nonumber
 &&\hspace{-10mm}=\frac{P({\rm REW}|M_{\rm UV,c})}{g'(M_{\rm UV,c})}\frac{{\rm REW}}{L_{\alpha}P(L_{\alpha})}P(M_{\rm UV,c})\\ 
&&\hspace{-10mm}=P({\rm REW}|M_{\rm UV,c})\frac{{\rm REW}}{L_{\alpha}P(L_{\alpha})} \frac{P(M_{\rm UV,c})}{g'(M_{\rm UV,c})},
\end{eqnarray}  
where in the last step we evaluated $g'(M_{\rm UV})$ at the pole of $g(M_{\rm UV})$, i.e. when $C \times {\rm REW} \times 10^{-0.4(M_{\rm UV,c}+s)}\equiv L_{\alpha}$. We can simplify this further to
\begin{equation}
P({\rm REW}|L_{\alpha})=-2.5 P({\rm REW}|M_{\rm UV,c})\frac{P(M_{\rm UV,c})}{\ln 10\hs L_{\alpha}P(L_{\alpha})}.
\end{equation} 
We finally compute the REW-PDF as
\begin{eqnarray}
 \nonumber
P({\rm REW})&=&\mathcal{N}\int_{L_{\alpha,{\rm min}}}^{L_{\alpha,{\rm max}}}dL_{\alpha}P({\rm REW}|L_{\alpha})P(L_{\alpha}) \\ 
 &&\hspace{-15mm}=\mathcal{N}\int_{L_{\alpha,{\rm min}}}^{L_{\alpha,{\rm max}}} P({\rm REW}|M_{\rm UV,c})P(M_{\rm UV,c})d\log_{10} L_{\alpha},
\end{eqnarray} where $\mathcal{N}$ is the normalization constant (which absorbed all numerical factors). 

\label{lastpage}
\end{document}